\documentclass[11pt,a4paper]{article}
\usepackage{times}
\usepackage{amssymb}
\usepackage{epsfig}
\usepackage{vmargin}
\usepackage{natbib}

\makeatletter \renewcommand{\@biblabel}[1]{#1.} \makeatother

\renewcommand{\cite}{\citep}

\begin{document}

\begin{bfseries}
\noindent\Large{\sc 
Labyrinthine Turing Pattern Formation \\ in the Cerebral Cortex}
\vspace{0.5cm}

\noindent\small{Julyan H. E. Cartwright}
\end{bfseries}

\begin{itshape}
\noindent\footnotesize
Laboratorio de Estudios Cristalogr\'aficos, CSIC, 
Facultad de Ciencias, E-18071 Granada, Spain. \\
E-mail julyan@lec.ugr.es, Web http://lec.ugr.es/$\sim$julyan

\end{itshape}

\normalsize

\vspace{0.2cm}


\vspace{0.2cm}

\noindent {\bf Published in:} J. Theor. Biol. {\bf 217}, 97--103, 2002. 

\vspace{0.2cm}

\begin{bfseries}
\noindent\small
I propose that the labyrinthine patterns of the cortices of mammalian brains 
may be formed by a Turing instability of interacting axonal guidance species
acting together with the mechanical strain imposed by the interconnecting axons.
\end{bfseries}

\section*{Introduction}

Labyrinthine patterns, of which those displayed on the surface of the mammalian
brain are a notable biological example, are some of the most striking
encountered in nature. A half-century ago Turing \cite{turing}, who was
studying morphogenesis in biological systems, had the insight that diffusion is
not always a homogenizing influence, and that spatial patterns would be formed
by a reaction--diffusion system combining local activation with long-range
inhibition. Spots and stripes on animal coats \cite{bard,murray}, patterning on
sea shells \cite{meinhardt}, stripes on tropical fish \cite{kondo}, and even
alligator teeth \cite{murray2}, have all since been put forward as examples of
Turing patterns in nature \cite{koch}. Here I propose that the morphogenesis of
labyrinthine patterns in the mammalian brain may likewise be an instance of
biological Turing pattern formation.

\begin{figure}[p]
\begin{center}
\def\epsfsize#1#2{0.5\columnwidth}
\leavevmode
\epsffile{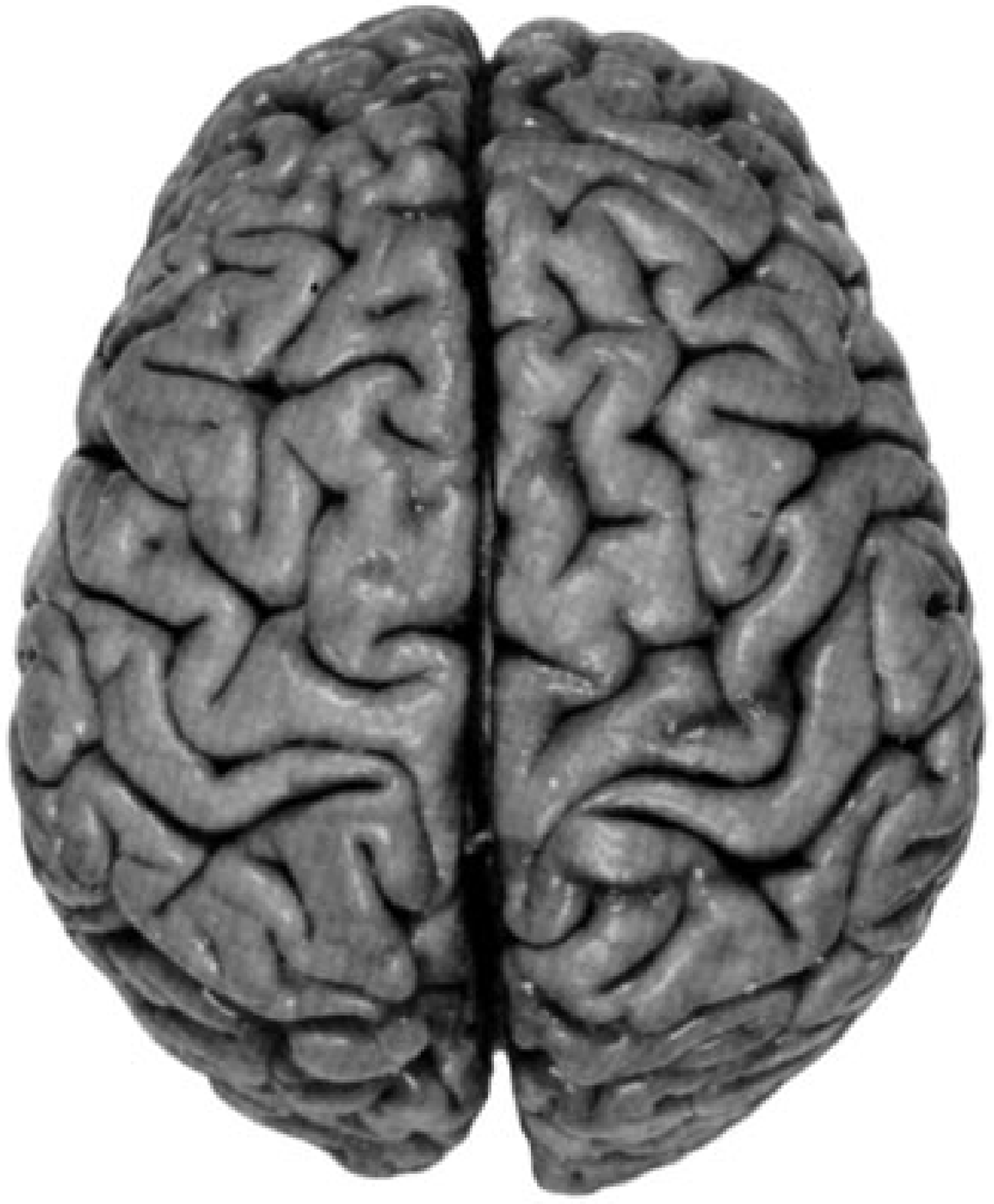}
\epsffile{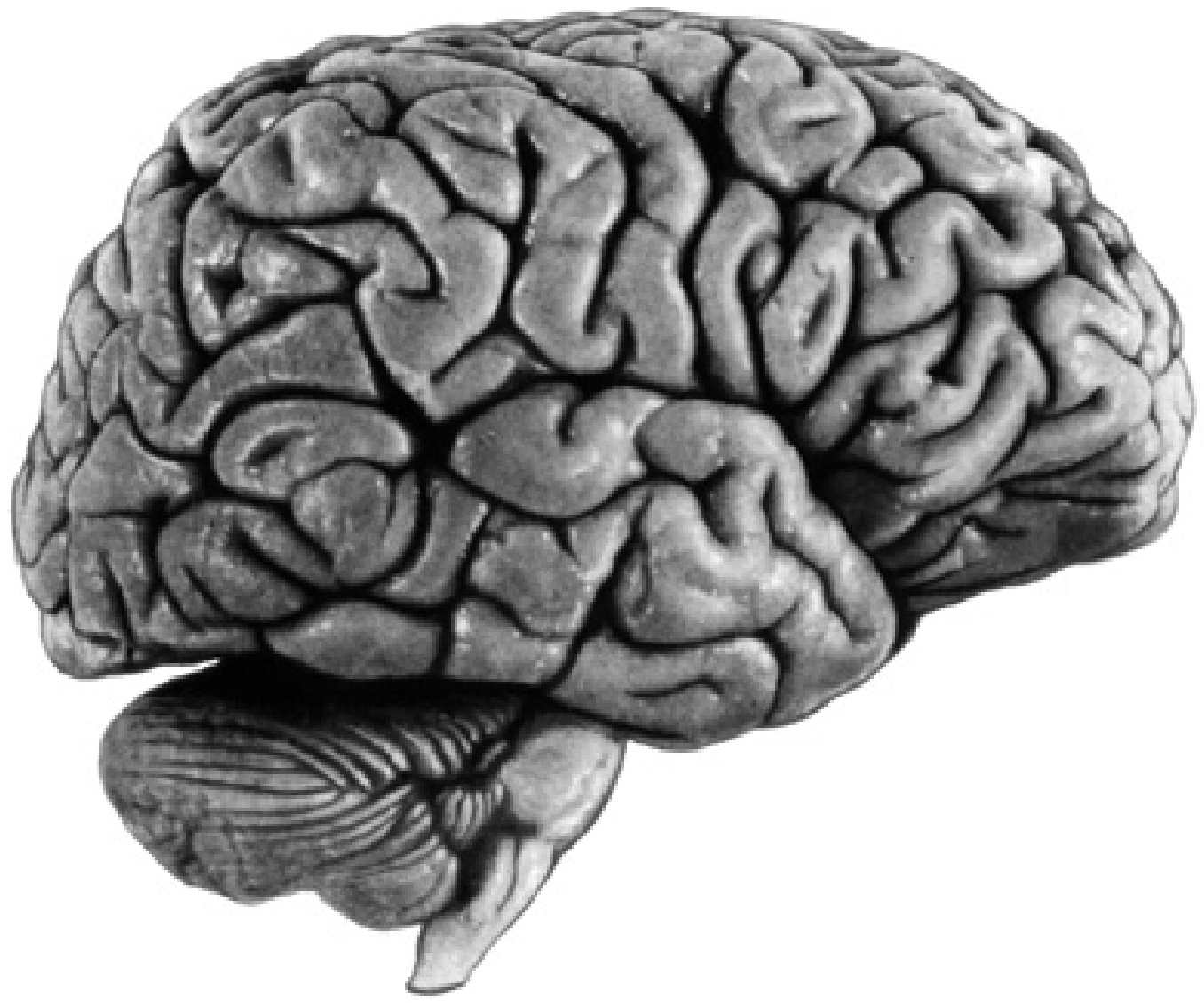}
\end{center}
\caption{\label{brain}
The labyrinthine patterns of the cerebral cortex of a human brain: (a) dorsal
view; (b) lateral view.
}
\end{figure}

The cerebral cortices of the higher mammals, and in particular humans, are made
up of a system of convoluted ridges (gyri) and valleys (sulci)
(Fig.~\ref{brain}). The pattern of gyri and sulci forms during the development
of the foetus: in humans the brain's surface is almost smooth at twenty-four
weeks, but between then and birth the cerebral cortex gradually assumes its
characteristic wrinkled form. This is not merely buckling, but is the result of
a deformation of the initial geometry, so that it would be necessary to
introduce cuts to smooth it out. As with the colouring of animal pelts, while
the overall morphology for a particular species is fixed, the detailed pattern
is unique to each individual, indicating that the specific form is decided
during development of the embryo. It is clear that the convolutions increase
the surface area of the cortex compared to the smooth brains of lower animals.
This allows a large cortical area to fit inside the head, and in addition may
minimize wiring volume, and benefit the synchronization of neuronal firings
\cite{griffin}. Whatever may be the functional origin of this pattern, however,
there remains the interesting question of its morphogenesis.

During embryonic development, neurons migrate from the thalamic areas within
the brain where they are produced to the cortex, where they wire themselves up
to other neurons via connecting axons \cite{rakic}. Over a century ago Ram\'on
y Cajal speculated \cite{ramon} that axons use chemical signals to find their
way. The growth cone at the tip of the axon senses gradients in the
concentration of axonal guidance species that may be either fixed to the
cellular substrate, or diffusing through it \cite{goodhill}. Many such
diffusive axonal guidance molecules have since been identified, some of which
are attractive (chemoattractants) or activate axon growth, while others are
repulsive (chemorepellants) or inhibit growth \cite{baier2,tessier}. On the
other hand, mechanical models for cortical folding have been put forward
\cite{richman,todd,raghavan}. It has recently been proposed \cite{vanessen}
that mechanical tension is the driving force for the formation of gyri and
sulci; that the surface of the cortex buckles and deforms under the tensile
loading imposed by the interconnecting axons. The idea has been shown to be
consistent with data on neuronal connectivity in the cat and the macaque
\cite{scannell}, and does a good job of explaining the presence of gyri and
sulci as high and low connectivity regions, but leaves open the question of the
overall form of the pattern: why is it labyrinthine? Here I marry the
mechanical-tension hypothesis to the idea of axonal pathfinding with diffusing
chemicals and show that the two together can provide an explanation for the
labyrinthine nature of the convolutions.

\section*{Reaction--diffusion equations and Turing patterns}

To model the pattern formation I propose an interaction between two diffusing
biologically active chemical species. One of the
species inhibits axon growth, and the other activates it, and moreover they
interact to activate and inhibit each other. A process of this form we may
describe mathematically by a pair of coupled reaction--diffusion equations
\begin{eqnarray}
&&\dot u=F(u,v)+\nabla^2 u, \nonumber \\
&&\dot v=G(u,v)+\delta\nabla^2 v
,\label{react-dif}\end{eqnarray}
where $u$ and $v$ are the scaled concentrations of the two diffusing and
reacting chemical species, with $u$ being the activator and $v$ the inhibitor.
$F(u,v)$ and $G(u,v)$ represent the interaction kinetics. Turing showed that a
spatially homogeneous steady state of Eq.\ (\ref{react-dif}) may become
unstable via a symmetry-breaking bifurcation if $\delta>1$, that is when the
inhibitor diffuses more rapidly than the activator. Associated with the
instability is a critical wave number that determines the wave length of the
resulting pattern. Numerical simulations of pattern formation in systems with
Turing instabilities have provided examples of the formation of patterns that
are periodic in one and two spatial dimensions (spots and stripes) \cite{judd},
together latterly with nonperiodic patterns including labyrinths
\cite{muratov2}. Moreover, a variety of chemical Turing patterns, including
stripes, spots, and rhombs, have been obtained in laboratory experiments
\cite{castets,ouyang,watzl}.

\begin{figure}[p]
\begin{center}
\def\epsfsize#1#2{0.7\columnwidth}
\leavevmode
\epsffile{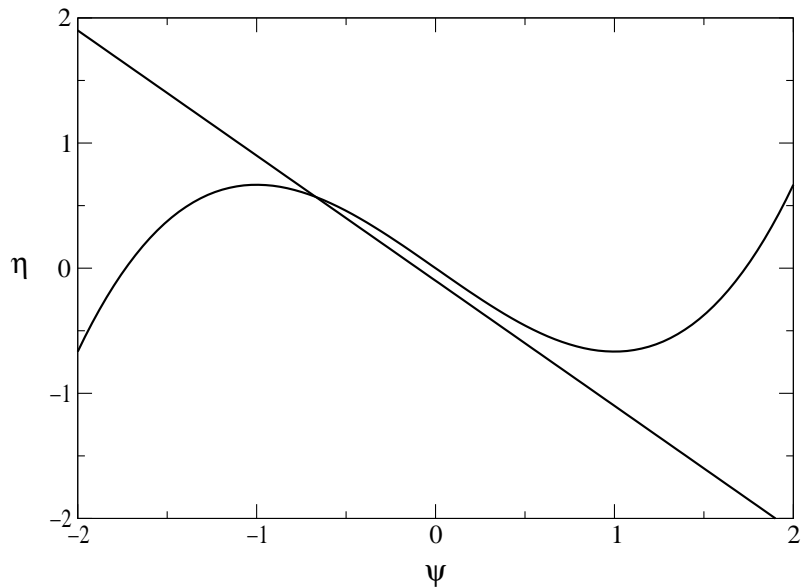}
\end{center}
\caption{\label{nullclines}
The system without diffusion; nullclines of (\ref{eq}) given by $\dot\psi=0$
and $\dot\eta=0$, that is $\eta=\psi^3/3-\psi$ and $\eta=-(\psi+\nu)/\beta$,
for the parameter values $\beta=1$, $\nu=0.1$ used in the simulations presented
here. For these parameter values there is a single equilibrium state, where the
nullclines cross, which is unstable, leading to pattern formation. In
many systems in which patterns form the nullclines have this characteristic 
N-shape \cite{quakeletter}.
}
\end{figure}

\begin{figure}[p]
\begin{center}
\def\epsfsize#1#2{\columnwidth}
\leavevmode
\epsffile{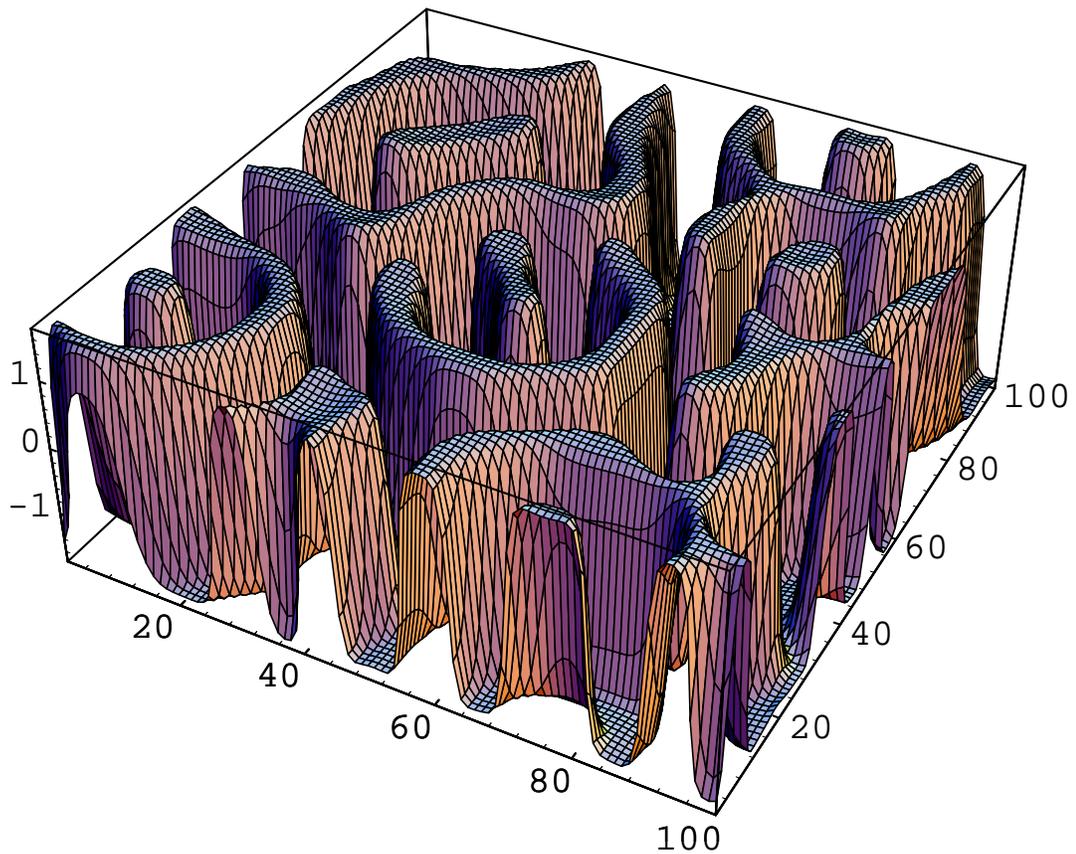}
\end{center}
\caption{\label{model}
Equations (\ref{eq}) integrated on a two-dimensional periodic domain with 
parameters $\gamma=2$, $\beta=1$, $\nu=0.1$, and $\delta=20$. Height represents
the concentration of activator. The static labyrinthine pattern is fully
formed at the time $t=40$ shown here and remains stable thereafter.
Concentration, time, and distance are in dimensionless units here.}
\end{figure}

Stripes and spots in biological settings have been explained with the Turing
bifurcation mechanism. The aim here, given that a Turing instability can
produce labyrinthine patterns, is to apply this understanding in a biological
context to explain their appearance in the brain. We as yet have no knowledge
of the molecular biology involved in the interactions. However, the
universality of the Turing bifurcation mechanism allows us to choose a simple
model for the activation and inhibition kinetics $F(u,v)$ and $G(u,v)$ of the
chemical species involved, knowing that the same qualitative effects can be
obtained with a more detailed mechanism faithful to the molecular biology. To
illustrate the interaction dynamics, I take the van der Pol--FitzHugh--Nagumo
equations \cite{vdp,fitz1,fitz2,nagumo} with diffusive coupling of both
variables
\begin{eqnarray}
\dot\psi&=&\gamma(\eta-\psi^3/3+\psi)+\nabla^2\psi, 
\nonumber \\
\dot\eta&=&-\gamma^{-1}(\psi+\nu+\beta\eta)+\delta\nabla^2\eta
\label{eq}
.\end{eqnarray}
Here $\psi$ represents the concentration of activator, and $\eta$ the
concentration of inhibitor, of an autocatalytic interaction whose kinetics are
determined by $\gamma$: for $\gamma<1$ local inhibition dominates, while for
$\gamma>1$, local activation is uppermost. $\nu$ and $\beta$ determine the
number and type of equilibrium states of the local dynamics; see
Fig.~\ref{nullclines}. The spatial interaction is determined by $\delta$: for
$\delta<1$ the inhibitor has shorter range than the activator, but for
$\delta>1$ the opposite is true. This van der Pol--FitzHugh--Nagumo model has
a homogeneous equilibrium state that is unstable to spatial oscillations via a
Turing bifurcation for $|\nu|<\sqrt{\delta(\delta\gamma^2-\beta)}\,
(3\delta\gamma^2-2\delta\gamma^2\beta-\beta^2)/(3\delta^2\gamma^3)$, and in
addition is unstable to temporal oscillations via a Hopf bifurcation for 
$|\nu|<\sqrt{\gamma^2-\beta}\,(3\gamma^2-2\gamma^2\beta-\beta^2)/(3\gamma^3)$.
Figure~\ref{model} presents the van der Pol--FitzHugh--Nagumo equations
integrated on a two-dimensional domain that represents the cerebral cortex. At
the parameter values of Fig.~\ref{model} a homogeneous initial condition is
unstable to both a Turing and a Hopf bifurcation, and a static labyrinthine
pattern is formed. The initial conditions contained a small amount of noise,
mimicking the variability of individuals, so in each simulation run the details
of the labyrinthine pattern were distinct, as the pattern developed following
the individual distributions of activator and inhibitor. The vertical axis
illustrates with the concentration of activator how mechanical tension
decorates the labyrinthine pattern with a three-dimensional structure of
convolutions, as axons growing preferentially in the activated regions force
these to grow upwards into heavily interconnected gyri, while the nonactivated
regions become the more sparcely connected sulci between them.

\section*{Biochemical and mechanochemical mechanisms}

One can conceive of several possible mechanisms for the interaction of axonal
guidance species with each other and with axons to produce the above dynamics.
The first possibility is that a direct chemical reaction between the species is
involved, as in chemical Turing pattern formation. This seems unlikely.  It
appears more probable that the interaction between activator and inhibitor
species is mediated by the medium, that is by cortical cells or axons. This
mechanism implies that cortical cells or axons would not merely passively 
sample concentration gradients of guidance molecules, but would actively alter
them. Such interactions have been postulated as necessary in diffusive axonal
guidance \cite{hentschel}. Furthermore, secretion of axonal guidance species by
axons \cite{zheng}, and interactions between molecules involved in axonal
guidance and growth \cite{ernst} have been observed, so it is plausible that an
activating and an inhibiting guidance species may be interacting at a cellular
level. In any case, the simple scenario presented above may be a caricature of
a more complicated series of interactions between more than two axonal guidance 
molecules. But the outcome of the process will nevertheless depend on the 
balance between activation and inhibition at each point, so the Turing 
bifurcation mechanism mediated by local activation plus long-range inhibition 
will still apply.

There is then the question of the timing of the laying down of the pattern.  Is
it concurrent with the axonal growth, so that the pattern appears as it forms,
or is it put into place earlier --- when the foetus is smaller --- as a
prepattern that the axons use for guidance later in embryonic development? Both
scenarios have been proposed in other biological Turing patterns:
prepatterning in animal coat markings \cite{bard,murray}, and concurrent
patterning in fish skin patterns \cite{kondo}. Here, developmental disorders
that are thought to have a predominantly genetic basis and that result in
malformation of the cerebral cortex \cite{dobyns} --- all of which lead to
severe mental retardation --- may give some insight into the mechanism of
pattern formation. The best known of these is lissencephaly, in which the
surface of the cortex does not have the normal gyri and sulci, and the brain is
smooth. Other conditions are pachygyria, in which the convoluted pattern is
larger-scaled than normal, with few, large gyri and sulci, and polymicrogyria,
which is the opposite case of the pattern being smaller-scaled than is usual,
so that there are many small sulci and gyri. Finally, so-called cobblestone
lissencephaly has been described, in which the brain has a cobbled or pebbled
surface. One can immediately see that these developmental errors are consistent
with a Turing instability. In pachygyria and polymicrogyria the wave length of
the pattern is respectively greater and less than usual, which could indicate
either, with concurrent patterning, that the ratio of diffusion coefficients
$\delta$ has altered or, in the case of the prepatterning scenario, that the
prepattern was laid down earlier and later than normal, respectively. On the
other hand, the cobblestone pattern may be interpreted as a change in the
morphology from a labyrinth to spots. This change can be produced in the  van
der Pol--FitzHugh--Nagumo model of Eqs (\ref{eq}) if the local kinetics are
altered so that the system remains unstable with respect to the Turing
instability but is stable to the Hopf instability \cite{muratov2}. In the same
way, lissencephaly may be caused by a change in the interaction dynamics
leading to the homogeneous state becoming stable to both the Hopf and Turing
instabilities, or alternatively may simply indicate the absence of the normal
axonal guidance species.

A further possible mechanism for the production of the convolutions is that
there is just one activating diffusive guidance molecule involved, and the part
of the other, inhibiting species is taken by the growing axons themselves. The
mechanical strain in the cortical tissue induced by the interconnecting axons
may then play the r\^ole of the long-range inhibitor.  The mathematics for this
mechanochemical model differ in detail from the reaction--diffusion system
presented in Eqs\ (\ref{react-dif}) and (\ref{eq}) above. However, this type of
model is closely related to the reaction--diffusion type --- the key concept of
long-range inhibition plus short-range activation being common to both --- and
is capable of forming similar patterns \cite{kauffman}. Mechanochemical models
have been proposed to model biological morphogenesis in systems such as
cellular morphology, in which mechanical strain in the cytoskeleton interacts
with the concentration of free calcium \cite{odell,oster,goodwin}. A more
detailed analysis of biologically faithful mechanochemical equations for the
formation  of sulci and gyri is underway; but I wish to emphasize that the
physics of the reaction--diffusion equations, presented above, to a large extent
carries over to the mechanochemical case, and allows the formation of similar
patterns. To distinguish between the biochemical and mechanochemical hypotheses,
and indeed between the prepatterning and concurrent patterning hypotheses,
will require extending our understanding of the molecular biology of axonal
guidance and growth.

\section*{Discussion}

Neuroanatomy classifies sulci and gyri as primary, secondary, or tertiary,
depending both upon their order of appearance in the developing brain, and upon
their variability between individuals. During cortical development, nuclei 
within the inner thalamic areas of the brain are first wired up to nearby
areas of the cortex; visual thalamic areas project to visual cortical areas,
auditory thalamic areas project to auditory cortical areas, and so on. These
cortical areas are subsequently interconnected during the process of forming
the convolutions of the cortex. In humans there are more than fifty functional
subdivisions of the cortex, but only in a few instances is there a correlation
between these areal boundaries and those of the convolutions \cite{zilles}.
This may well be because neighbouring areas may have
connections along their mutual boundary comparable to the connectivity
within each area alone, as folding {\em is} correlated with connectivity
\cite{scannell}. If the initial conditions for the cortical pattern formation
were completely uniform, perturbed only by noise, as in the simulation of
Fig.~\ref{model}, then the convolutions for each individual would emerge
completely uncorrelated. In fact, differences in the folding pattern between
individuals increase from primary to secondary to tertiary convolutions; the
primary convolutions are more or less fixed in place, while it is at the level
of the tertiary convolutions that there is freedom for the pattern to develop
in a much more variable manner. An example of this variability is provided by
Einstein's brain, examination of which has shown the absence of a sulcus found
in other subjects in the inferior parietal area; it has been surmised that the
consequent greater neuronal connectivity in that region, which is responsible
for mathematical thought, may help to account for his genius \cite{witelson}. 
Furthermore, abnormal sensory experience can lead to abnormal convolutions
\cite{rakic}; the probable chain of events is that  sensory experience
influences connectivity, and thence the folding pattern. All this implies that
the initial condition for pattern formation contains information at the level
of the primary convolutions. The thalamic input determines the initial
conditions for the folding pattern through the pattern of cortical projections.
The Turing mechanism takes these initial conditions and moulds from them a
labyrinthine pattern of convolutions through connection-density-dependent
growth and mechanical tension. The physical Turing mechanism may then be
operating at the level of the tertiary convolutions, under genetic control at
the level of the thalamic input.

The Turing instability produces a labyrinthine pattern from an initially
homogeneous system by symmetry breaking. In physics and chemistry --- in
excitable chemical reactions, block copolymers, magnetic fluids, and
superconductors \cite{dickstein,seul}  --- labyrinthine patterns have been
explained as the result of local activation combined with long-range inhibition
\cite{goldstein2}. In particular, the spontaneous appearance of labyrinths in a
chemical reaction has been seen as a Turing instability in a
reaction--diffusion system. I have set out here to understand biological
labyrinthine patterns in this context. I have shown that labyrinthine
patterning in the cerebral cortex is consistent with a Turing instability. This
mechanism joins together two familiar ideas in cortical development: the part
played by axonal pathfinding chemicals in guiding neuronal connectivity, and
the role of mechanical tension in the uplifting of cortical areas into gyri.
The hypothesis that diffusing and interacting axonal guidance molecules cause a
Turing instability allows one to understand the breakdown of homogeneity during
cortical development; at the same time the mechanical tension induced by the
high connectivity in activated regions explains why the gyri and sulci form a
labyrinth. The idea may provide valuable insight into developmental disorders
of the cortex like lissencephaly. There is as yet no molecular-biological
evidence for such a Turing mechanism, but I hope that the present article will
stimulate such research. The present hypothesis is not meant to suggest that
genes have no r\^ole in the production of the convolutions of the cerebral
cortex, but rather to propose that this biological process may take advantage
of the physical mechanism of pattern formation induced by the mathematical
theory of Turing bifurcation.

\section*{Acknowledgements}

I am indebted to Jack Scannell for insightful comments that have improved this
work. I should like to thank the Wellcome Department of Cognitive Neurology,
London, UK for Fig.~\ref{brain}, and Miguel Hern\'andez University, Elche,
Spain, for my time there in 1998 during which I conceived these ideas. I
acknowledge the financial support of the Spanish Consejo Superior de
Investigaciones Cient\'{\i}ficas.

\section*{References}

\bibliographystyle{bifchaos}
\bibliography{database}

\begin{thebibliography}{}

\bibitem[\protect\citeauthoryear{Baier \& Bonhoeffer}{1994}]{baier2}
Baier, H. \& Bonhoeffer, F. [1994]
\newblock ``Attractive axon guidance molecules,''
\newblock {\em Science} {\bf 265}, 1541--1542.

\bibitem[\protect\citeauthoryear{Bard}{1981}]{bard}
Bard, J. B.~L. [1981]
\newblock ``A model for generating aspects of zebra and other mammalian coat
  patterns,''
\newblock {\em J. Theor. Biol.} {\bf 93}, 363--385.

\bibitem[\protect\citeauthoryear{Cartwright \bgroup \em et al.\egroup
  }{1997}]{quakeletter}
Cartwright, J. H.~E., Hern\'andez-Garc\'{\i}a, E. \& Piro, O. [1997]
\newblock ``{Burridge-Knopoff} models as elastic excitable media,''
\newblock {\em Phys. Rev. Lett.} {\bf 79}, 527--530.

\bibitem[\protect\citeauthoryear{Castets \bgroup \em et al.\egroup
  }{1990}]{castets}
Castets, V., Dulos, E., Boissonade, J. \& de~Kepper, P. [1990]
\newblock ``Experimental evidence of a sustained standing {Turing}-type
  nonequilibrium chemical pattern,''
\newblock {\em Phys. Rev. Lett.} {\bf 64}, 2953--2956.

\bibitem[\protect\citeauthoryear{Dickstein \bgroup \em et al.\egroup
  }{1993}]{dickstein}
Dickstein, A.~J., Erramilli, S., Goldstein, R.~E., Jackson, D.~P. \& Langer,
  S.~A. [1993]
\newblock ``Labyrinthine pattern formation in magnetic fluids,''
\newblock {\em Science} {\bf 261}, 1012--1015.

\bibitem[\protect\citeauthoryear{Dobyns \& Truwit}{1995}]{dobyns}
Dobyns, W.~B. \& Truwit, C.~L. [1995]
\newblock ``Lissencephaly and other genetic neuronal migration disorders: 1995
  update,''
\newblock {\em Neuropediatrics} {\bf 26}, 132--147.

\bibitem[\protect\citeauthoryear{Ernst \bgroup \em et al.\egroup
  }{2000}]{ernst}
Ernst, A.~F., Gallo, G., Letorneau, P.~C. \& McLoon, S.~C. [2000]
\newblock ``Stabilization of growing retinal axons by the combined signaling of
  nitric oxide and brain-derived neurotropic factor,''
\newblock {\em J. Neurosci.} {\bf 20}, 1458--1469.

\bibitem[\protect\citeauthoryear{FitzHugh}{1960}]{fitz1}
FitzHugh, R.~A. [1960]
\newblock ``Thresholds and plateaus in the {Hodgkin--Huxley} nerve equations,''
\newblock {\em J. Gen. Physiol.} {\bf 43}, 867--896.

\bibitem[\protect\citeauthoryear{FitzHugh}{1961}]{fitz2}
FitzHugh, R.~A. [1961]
\newblock ``Impulses and physiological states in theoretical models of nerve
  membrane,''
\newblock {\em Biophys. J.} {\bf 1}, 445--466.

\bibitem[\protect\citeauthoryear{Goldstein \bgroup \em et al.\egroup
  }{1996}]{goldstein2}
Goldstein, R.~E., Muraki, D.~J. \& Petrich, D.~M. [1996]
\newblock ``Interface proliferation and the growth of labyrinths in a
  reaction-diffusion system,''
\newblock {\em Phys. Rev. E} {\bf 53}, 3933--3957.

\bibitem[\protect\citeauthoryear{Goodhill \& Urbach}{1999}]{goodhill}
Goodhill, G.~J. \& Urbach, J.~S. [1999]
\newblock ``Theoretical analysis of gradient detection by growth cones,''
\newblock {\em J. Neurobiol.} {\bf 41}, 230--241.

\bibitem[\protect\citeauthoryear{Goodwin \& Trainor}{1985}]{goodwin}
Goodwin, B.~C. \& Trainor, L. E.~H. [1985]
\newblock ``Tip and whorl morphogenesis in {\em {a}cetabularia} by
  calcium-related strain fields,''
\newblock {\em J. Theor. Biol.} {\bf 117}, 79--106.

\bibitem[\protect\citeauthoryear{Griffin}{1994}]{griffin}
Griffin, L.~D. [1994]
\newblock ``The intrinsic geometry of the cerebral cortex,''
\newblock {\em J. Theor. Biol.} {\bf 166}, 261--273.

\bibitem[\protect\citeauthoryear{Hentschel \& van Ooyen}{1999}]{hentschel}
Hentschel, H. G.~E. \& van Ooyen, A. [1999]
\newblock ``Models of axon guidance and bundling during development,''
\newblock {\em Proc. Roy. Soc. Lond. B} {\bf 266}, 2231--2238.

\bibitem[\protect\citeauthoryear{Judd \& Silber}{2000}]{judd}
Judd, S.~L. \& Silber, M. [2000]
\newblock ``Simple and superlattice {Turing} patterns in reaction--diffusion
  systems: Bifurcation, bistability, and parameter collapse,''
\newblock {\em Physica D} {\bf 136}, 45--65.

\bibitem[\protect\citeauthoryear{Kauffman}{1993}]{kauffman}
Kauffman, S.~A. [1993]
\newblock {\em The Origins of Order}
\newblock (Oxford University Press).

\bibitem[\protect\citeauthoryear{Koch \& Meinhardt}{1994}]{koch}
Koch, A.~J. \& Meinhardt, H. [1994]
\newblock ``Biological pattern formation: From basic mechanisms to complex
  structures,''
\newblock {\em Rev. Mod. Phys.} {\bf 66}, 1481--1507.

\bibitem[\protect\citeauthoryear{Kondo \& Asai}{1995}]{kondo}
Kondo, S. \& Asai, R. [1995]
\newblock ``A reaction--diffusion wave on the skin of the marine angelfish {\em
  pomacanthus},''
\newblock {\em Nature} {\bf 376}, 765--768.

\bibitem[\protect\citeauthoryear{Meinhardt}{1997}]{meinhardt}
Meinhardt, H. [1997]
\newblock ``Biological pattern formation as a complex dynamic phenomenon,''
\newblock {\em Int. J. Bifurcation \& Chaos} {\bf 7}, 1--26.

\bibitem[\protect\citeauthoryear{Muratov \& Osipov}{1996}]{muratov2}
Muratov, C.~B. \& Osipov, V.~V. [1996]
\newblock ``Scenarios of domain pattern formation in a reaction--diffusion
  system,''
\newblock {\em Phys. Rev. E} {\bf 54}, 4860--4879.

\bibitem[\protect\citeauthoryear{Murray}{1989}]{murray}
Murray, J.~D. [1989]
\newblock {\em Mathematical Biology}
\newblock (Springer).

\bibitem[\protect\citeauthoryear{Murray \& Kulesa}{1996}]{murray2}
Murray, J.~D. \& Kulesa, P.~M. [1996]
\newblock ``On a dynamic reaction--diffusion mechanism: The spatial patterning
  of teeth primordia in the alligator,''
\newblock {\em J. Chem. Soc., Faraday Trans.} {\bf 92}, 2927--2932.

\bibitem[\protect\citeauthoryear{Nagumo \bgroup \em et al.\egroup
  }{1962}]{nagumo}
Nagumo, J.~S., Arimoto, S. \& Yoshizawa, S. [1962]
\newblock ``An active pulse transmission line simulating nerve axon,''
\newblock {\em Proc. IREE Aust.} {\bf 50}, 2061--2070.

\bibitem[\protect\citeauthoryear{Odell \bgroup \em et al.\egroup
  }{1981}]{odell}
Odell, G., Oster, G., Burnside, B. \& Alberch, P. [1981]
\newblock ``The mechanical basis of morphogenesis. {I}: Epithelial folding and
  invagination,''
\newblock {\em Dev. Biol.} {\bf 85}, 446--462.

\bibitem[\protect\citeauthoryear{Oster \& Odell}{1984}]{oster}
Oster, G. \& Odell, G. [1984]
\newblock ``The mechanochemistry of cytogels,''
\newblock {\em Physica D} {\bf 12}, 333--350.

\bibitem[\protect\citeauthoryear{Ouyang \& Swinney}{1991}]{ouyang}
Ouyang, Q. \& Swinney, H.~L. [1991]
\newblock ``Transition from a uniform state to hexagonal and striped {Turing}
  patterns,''
\newblock {\em Nature} {\bf 352}, 610--611.

\bibitem[\protect\citeauthoryear{Raghavan \bgroup \em et al.\egroup
  }{1997}]{raghavan}
Raghavan, R., Lawton, W., Ranjan, S.~R. \& Viswanathan, R.~R. [1997]
\newblock ``A continuum mechanics-based model for cortical growth,''
\newblock {\em J. Theor. Biol.} {\bf 187}, 285--296.

\bibitem[\protect\citeauthoryear{Rakic}{1988}]{rakic}
Rakic, P. [1988]
\newblock ``Specification of cerebral cortical areas,''
\newblock {\em Science} {\bf 241}, 170--176.

\bibitem[\protect\citeauthoryear{{Ram\'on y Cajal}}{1893}]{ramon}
{Ram\'on y Cajal}, S. [1893]
\newblock ``La r\'etine des vertebres,''
\newblock {\em La Cellule} {\bf 9}, 119--258.

\bibitem[\protect\citeauthoryear{Richman \bgroup \em et al.\egroup
  }{1975}]{richman}
Richman, D.~R., Stewart, R.~M., Hutchinson, J.~W. \& Caviness, Jr., V.~S.
  [1975]
\newblock ``Mechanical model of brain convolutional development,''
\newblock {\em Science} {\bf 189}, 18--21.

\bibitem[\protect\citeauthoryear{Scannell}{1997}]{scannell}
Scannell, J.~W. [1997]
\newblock ``Determining cortical landscapes,''
\newblock {\em Nature} {\bf 386}, 452.

\bibitem[\protect\citeauthoryear{Seul \& Andelman}{1995}]{seul}
Seul, M. \& Andelman, D. [1995]
\newblock ``Domain shapes and patterns: The phenomenology of modulated
  phases,''
\newblock {\em Science} {\bf 267}, 476--483.

\bibitem[\protect\citeauthoryear{Tessier-Lavigne \& Goodman}{1996}]{tessier}
Tessier-Lavigne, M. \& Goodman, C.~S. [1996]
\newblock ``The molecular biology of axon guidance,''
\newblock {\em Science} {\bf 274}, 1123--1133.

\bibitem[\protect\citeauthoryear{Todd}{1982}]{todd}
Todd, P.~H. [1982]
\newblock ``A geometric model for the cortical folding pattern of simple folded
  brains,''
\newblock {\em J. Theor. Biol.} {\bf 97}, 529--538.

\bibitem[\protect\citeauthoryear{Turing}{1952}]{turing}
Turing, A.~M. [1952]
\newblock ``The chemical basis of morphogenesis,''
\newblock {\em Phil. Trans. Roy. Soc. Lond. B} {\bf 237}, 37--72.

\bibitem[\protect\citeauthoryear{van~der Pol \& van~der Mark}{1928}]{vdp}
van~der Pol, B. \& van~der Mark, J. [1928]
\newblock ``The heart beat considered as a relaxation oscillator and an
  electrical model of the heart,''
\newblock {\em Phil. Mag. (7)} {\bf 6}, 763--775.

\bibitem[\protect\citeauthoryear{van Essen}{1997}]{vanessen}
van Essen, D.~C. [1997]
\newblock ``A tension-based theory of morphogenesis and compact wiring in the
  central nervous system,''
\newblock {\em Nature} {\bf 385}, 313--318.

\bibitem[\protect\citeauthoryear{Watzl \& M{\"u}nster}{1995}]{watzl}
Watzl, M. \& M{\"u}nster, A.~F. [1995]
\newblock ``Turing-like spatial patterns in a polyacrylamide-methylene
  blue--sulfide--oxygen system,''
\newblock {\em Chem. Phys. Lett.} {\bf 242}, 273--278.

\bibitem[\protect\citeauthoryear{Witelson \bgroup \em et al.\egroup
  }{1999}]{witelson}
Witelson, S.~F., Kigar, D.~L. \& Harvey, T. [1999]
\newblock ``The exceptional brain of {Albert Einstein},''
\newblock {\em Lancet} {\bf 353}, 2149--2153.

\bibitem[\protect\citeauthoryear{Zheng \bgroup \em et al.\egroup
  }{1994}]{zheng}
Zheng, J.~Q., Felder, M., Connor, J.~A. \& Poo, M.~M. [1994]
\newblock ``Turning of nerve growth cones induced by neurotransmitters,''
\newblock {\em Nature} {\bf 368}, 140--144.

\bibitem[\protect\citeauthoryear{Zilles \bgroup \em et al.\egroup
  }{1997}]{zilles}
Zilles, K., Schleicher, A., Langemann, C., Amunts, K., Morosan, P.,
  Palomero-Gallagher, N., Schormann, T., Mohlberg, H., B{\"u}rgel, U.,
  Steinmetz, H., Schlaug, G. \& Roland, P.~E. [1997]
\newblock ``Quantitative analysis of sulci in the human cerebral cortex:
  Development, regional heterogeneity, gender difference, asymmetry,
  intersubject variability and cortical architecture,''
\newblock {\em Human Brain Mapping} {\bf 5}, 218--221.

\end{thebibliography}

\end{document}